**Orbital Current-Driven Magnetization Switching in a Magnetic Tunnel Junction**

Jingkai Xu[1+], Dongxing Zheng[1+], Meng Tang[1+], Chen Liu[1], Bin He[1], Man Yang[1], Hao Li[1], Yan Li[1], Aitian Chen[2], Senfu Zhang[3], Ziqiang Qiu[4], Xixiang Zhang[1*]

[1]Physical Science and Engineering Division, King Abdullah University of Science and Technology (KAUST), Thuwal 23955–6900, Saudi Arabia

[2]State Key Laboratory of Electronic Thin Film and Integrated Devices, School of Physics, University of Electronic Science and Technology of China, Chengdu, 611731 China

[3]Key Laboratory for Magnetism and Magnetic Materials of Ministry of Education, Lanzhou University, Lanzhou 730000, China

[4]Department of Physics, University of California at Berkeley, Berkeley, CA 94720, USA

[*]Corresponding authors:

Xixiang Zhang (email: xixiang.zhang@kaust.edu.sa)

[+]These authors contributed equally to this work.

**Abstract**

Spin-orbitronics, based on both spin and orbital angular momentum, presents a promising pathway for energy-efficient memory and logic devices. Recent studies have demonstrated the emergence of orbital currents in light transition metals such as Ti, Cr, and Zr, broadening the scope of torque-driven magnetization switching. In particular, the orbital Hall effect (OHE), which arises independently of spin-orbit coupling (SOC), has shown potential for enhancing torque efficiency in spintronic devices. However, the direct integration of orbital current into magnetic tunnel junction (MTJ) remains unexplored. In this work, we design a light metal/heavy metal/ferromagnet multilayer structure and experimentally demonstrate magnetization switching by orbital current. Furthermore, we have realized a robust spin-orbit torque MTJ cell by incorporating a reference layer that is pinned by a synthetic antiferromagnetic structure. We observed a tunnel magnetoresistance of 66%, evident in both magnetic field and current-driven switching processes. Our findings underscore the potential for employing orbital current in designing next-generation spintronic devices.

## Introduction

The magnetic tunnel junction (MTJ) is one of the most fundamental building blocks of magnetic random-access memory (MRAM), a next-generation non-volatile memory technology that offers significant advantages over conventional memory systems, including low power consumption, high endurance, and ultrafast operation.[1-3] Consisting of two ferromagnetic layers separated by a thin insulating barrier, data storage is achieved by controlling the relative magnetization orientation of these ferromagnetic layers, leading to a change in tunnel magnetoresistance (TMR), which can be detected to encode binary information ("0" and "1").[4-9] Now, a critical challenge in MRAM development is the efficient manipulation of the free-layer magnetization. However, this can be achieved by the MTJ in three different switching mechanisms: driven by the external magnetic field, spin-transfer torque (STT), and spin orbit torque (SOT). In the SOT mechanism, the writing current flows transversely in a separate bottom electrode rather than through the tunnel barrier, mitigating electromigration-induced degradation and significantly enhancing device endurance.[10, 11] Various methods have been explored to minimize switching current density ($J_c$) to improve the performance of the MTJ, including the utilization of materials with large spin Hall angles, such as topological insulators,[12-16] two-dimensional materials,[17, 18] and complex oxides.[19-23] While these materials exhibit high charge-to-spin conversion efficiency and can substantially reduce $J_c$, their limited compatibility with standard memory device fabrication processes limits their practical application in commercial MRAM technology.

Recently, the orbital Hall effect (OHE) has garnered significant attention as a promising mechanism for generating orbital currents in response to an applied electric field in a transverse direction, providing an efficient route to enhance torque efficiencies.[24-26] Unlike the conventional spin Hall effect, OHE does not rely on strong spin-orbit coupling (SOC), eliminating the need for rare and expensive heavy metals.[25-29] This is a major advantage as it allows for the use of lighter, more abundant, and potentially more cost-effective materials like Ti or Cu for spintronic and orbitronic applications. This broadens the material scope and flexibility in future electronics applications and reduces reliance on expensive heavy metals.[30-32] Furthermore, OHE exhibits a larger orbital Hall angle,[33, 34] which can significantly enhance torque efficiency.[35-41] Both theory and experiments have shown that the efficiency of orbital moments may exceed that of spin-orbit moments of traditional heavy metals. Unlike spin angular momentum, which can be given by $S = \sqrt{s(s+1)}\hbar = \frac{\sqrt{3}}{2}\hbar$, a key advantage of orbital moments is that the magnitude of an electron's orbital angular momentum is not an intrinsic, unchanging property. Instead, it depends on the azimuthal quantum number ($L = \sqrt{l(l+1)}\hbar$), offering a tunable parameter for designing more efficient spintronic devices. As a result, an electron's OAM can be much larger than its spin, potentially by orders of magnitude. In some topological insulators, for instance, the OHE can be orders of magnitude larger than the SHE.[42] Thereby, it can achieve magnetic moment reversal with lower power consumption and higher efficiency.[31, 32] Crucially, OHE-based materials are compatible with standard memory device fabrication processes, making them highly attractive for next-generation MRAM technologies.[31]

In this work, we demonstrate perpendicular magnetization switching in $Co_{20}Fe_{60}B_{20}$ (CFB) driven by an orbital current generated via the OHE in a Ti layer within a Ti/Ta/CFB/MgO/Ta multilayer. The orbital current generated by the Ti layer is converted into a spin current through SOC in the adjacent Ta layer, facilitating efficient magnetization switching.[43] Compared to the conventional Ta/CFB bilayer, the inclusion of the Ti layer reduces the critical switching current density. To explore its practical viability, we fabricated a perpendicular magnetic tunnel junction (pMTJ) incorporating the Ti/Ta/CFB free layer, achieving a TMR ratio of 66% at room temperature. These results highlight the future potential of OHE-driven magnetization for next-generation SOT-MRAM devices.

**Results and Discussion**

Figure 1a illustrates the design of our MTJ, which features two ferromagnetic (FM) layers separated by a thin insulating barrier. When a spin-polarized current is injected into the ferromagnetic layer, the electrons with specific spin orientations exert torques on the magnetic moments of the ferromagnets, inducing magnetization switching. In this work, we utilize the orbital current generated by the OHE in a light metal layer to drive magnetization switching. Since orbital current cannot directly interact with the spin-based magnetic moments,[43-47] a heavy metal with strong SOC is introduced between the light metal and the ferromagnetic layer. This heavy metal layer facilitates the conversion of orbital currents into spin currents, which then exerts SOT on the ferromagnetic layer. To achieve this goal, we designed a multilayer stack consisting of Ti/Ta/CFB/MgO/CFB/synthetic antiferromagnet, as shown in Fig. 1b. The CFB is selected as the ferromagnetic layers, owing to its large spin polarization (~65%) and perpendicular magnetic anisotropy (PMA).[48, 49] A synthetic antiferromagnetic (SAF) stack is placed atop the CFB layers to pin the magnetization of the top CFB layer.[50-52] Directly beneath the ferromagnetic stack lies a thin Ta layer, a heavy metal with strong SOC, which not only facilitates spin conversion but also enhances the PMA of CFB.[49] Below the Ta layer, the Ti layer—despite its weak intrinsic SOC—generates an orbital current through the OHE. This orbital current is converted into spin currents by the Ta layer, effectively inducing magnetization by switching in the CFB free layer.

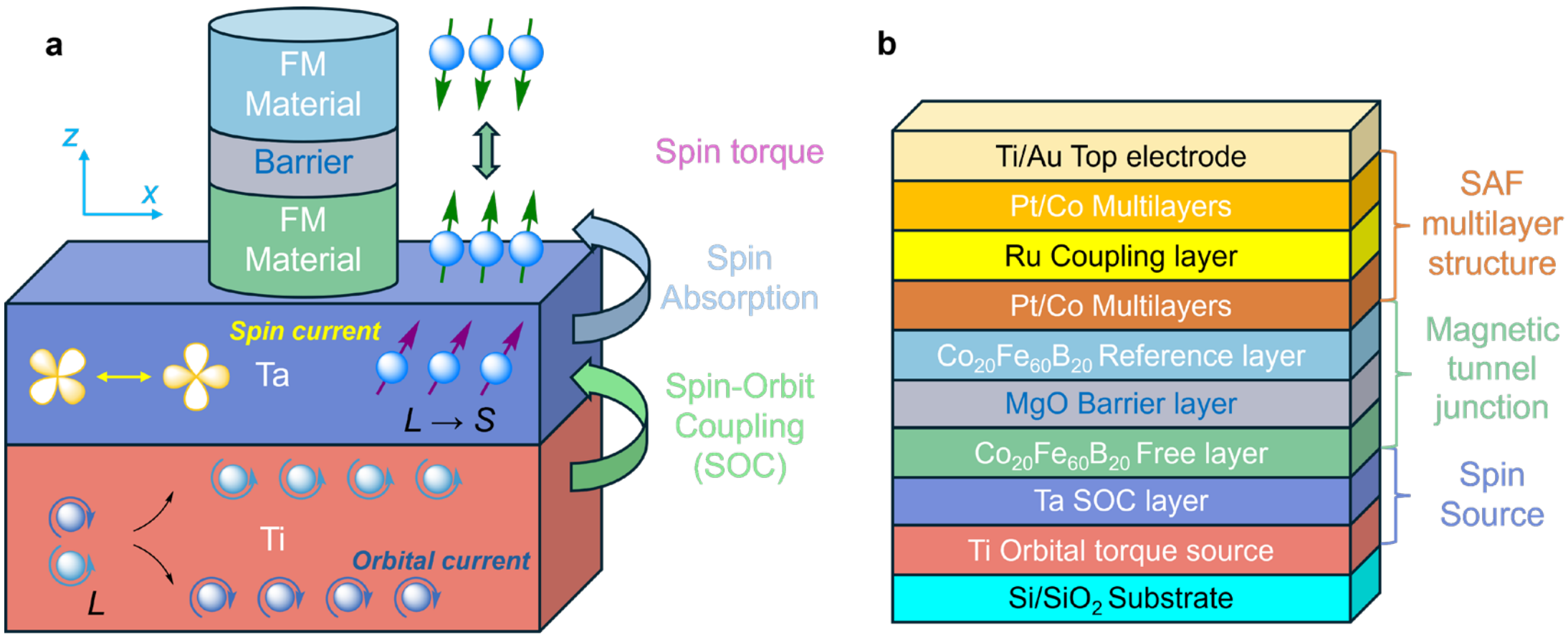

**Figure 1.** (a) Schematic of the SOT-MRAM cell design. The orbital current is converted into a spin current, which then induces the magnetization switching in the free layer. (b) Schematic of the multilayer stack design for the orbital current-driven pMTJs.

### Spin transport measurements

Based on this design, we first aimed to achieve a CFB layer with PMA above the Ti/Ta layers. We deposited a series of Ti(5.0)/Ta(0.5)/CFB($t$)/MgO(2.2)/Ta(2.0) stacks on the oxidized Si(100) wafers using a Singulus ROTARIS magnetron sputtering system, with all thicknesses in nanometers. A 2.0 nm Ta layer is deposited on the top, serving as a capping layer to protect the multilayer from oxidation. The magnetic properties are characterized via magneto-optical Kerr effect (MOKE) measurements. PMA is observed in the samples with CFB thicknesses ranging from 0.75 nm to 0.85 nm, as evidenced by sharp magnetization switching (see Fig. S3 in the supporting information). Among these, the CFB layer with $t$ = 0.75 nm is selected for its higher coercive field of ~2.5 mT (Fig. 2a), which enhances the magnetic stability—crucial for robust data storage and processing. Furthermore, magnetic-domain images captured during magnetic field-driven magnetization switching measurement confirm a domain-wall motion mechanism.

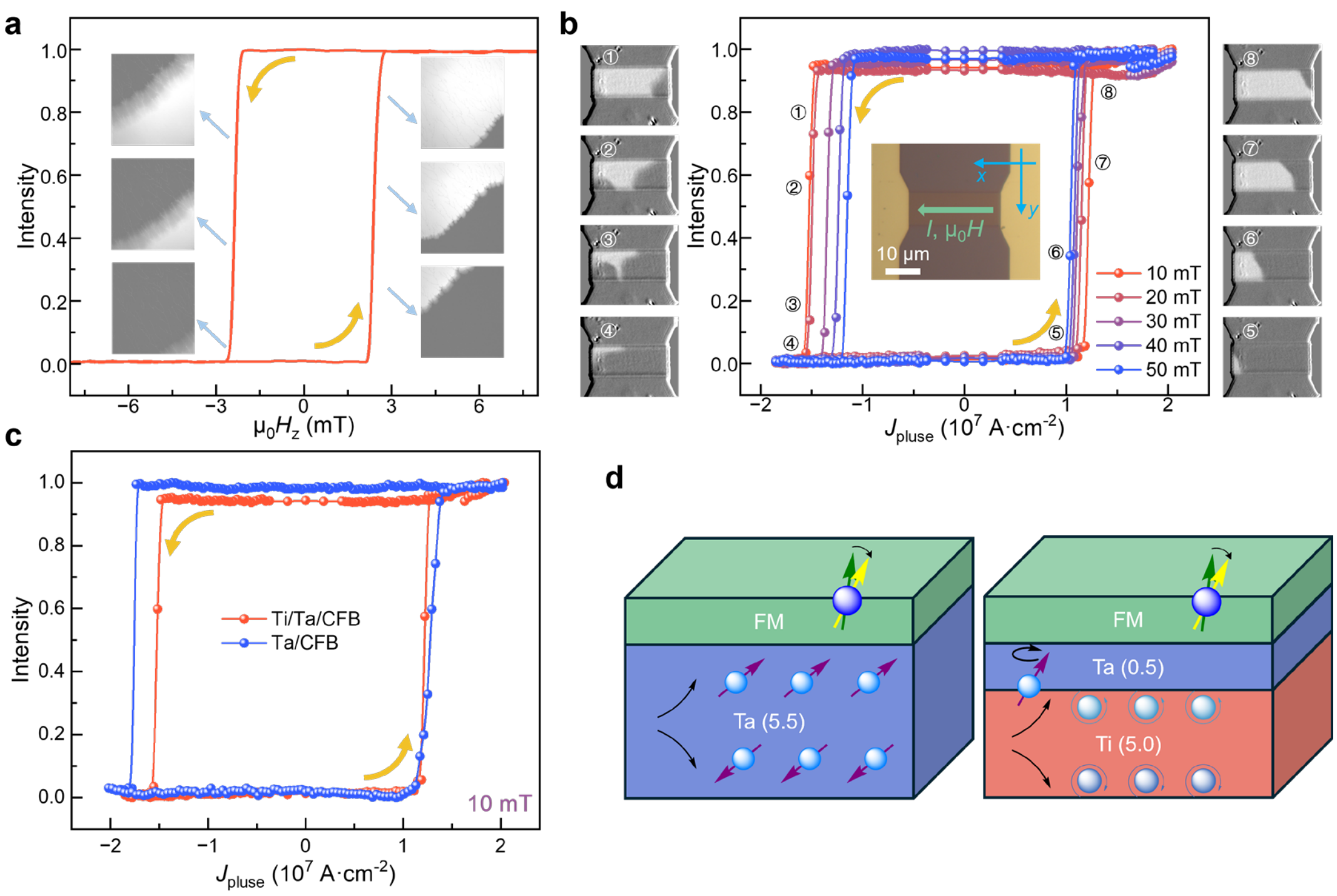

**Figure 2.** (a) Magnetization switching measurement using polar MOKE in the Ti(5)/Ta(0.5)/CFB(0.75)/MgO(2.2)/Ta(2.0) multilayer. The intensity of polarized light passing through the polarizer is recorded as a function of the magnetic field $\mu_0 H_z$, with corresponding MOKE images at different $\mu_0 H_z$ values. (b) Current-induced magnetization switching as a function of pulse current under varying in-plane magnetic fields $\mu_0 H_x$, with a pulse width of 100 us. The corresponding MOKE images at different pulse currents are also shown. The inset displays a microscope image of the device used for current-driven magnetization switching. (c) Current-induced magnetization switching curves for Ti(5)/Ta(0.5)/CFB(0.75)/MgO(2.2)/Ta(2.0) and Ta(5.5)/CFB(0.75)/MgO(2.2)/Ta(2.0) multilayers. (d) Schematic illustration of the two different switching mechanisms in the multilayers with and without the Ti layer.

After confirming PMA in the Ti(5.0)/Ta(0.5)/CFB(0.75)/MgO(2.2)/Ta(2.0) multilayers, we performed the current-induced magnetization switching measurements. Fig. 2b shows the results of these measurements. During the experiment, an in-plane magnetic field ($\mu_0 H_x$) is applied along the current direction to facilitate deterministic magnetization switching. The MOKE signal reveals that the magnetization is switched by the pulse current: a positive pulse favors higher MOKE intensity, while a negative pulse favors lower MOKE intensity. Due to the stripe structure of the stack, the observed switching mechanism is dominated by domain-wall motion rather than coherent rotation switching, as indicated by the corresponding MOKE images in Fig. 2b. Furthermore, the critical switching current density ($J_c$) is found to depend on the external magnetic field $\mu_0 H_x$, increasing as $\mu_0 H_x$ decreases (See page 4 and page 5 in the Supporting information for the detailed method to calculate the current density in the multilayers). This trend is consistent with previous observations that $J_C$ is influenced by the applied external magnetic field.[53]

After verifying the current induced magnetization switching in the Ti(5)/Ta(0.5)/CFB(0.75)/MgO(2.2)/Ta(2.0) multilayer, we conducted a comparative analysis of the current induced magnetization switching measurement between the Ti(5)/Ta(0.5)/CFB(0.75)/MgO(2.2)/Ta(2.0) structure and the Ta(5.5)/CFB(0.75)/ MgO(2.2)/Ta(2.0) structure under a magnetic field of $\mu_0 H_x$ = 10 mT, as shown in Fig. 2c. The $J_c$ for Ti(5)/Ta(0.5)/CFB(0.75)/MgO(2.2)/Ta(2.0) multilayer is ~$1.2\times10^7$ A/cm$^2$, lower than ~$1.5\times10^7$ A/cm$^2$ for the Ta(5.5)/ CFB(0.75)/MgO(2.2)/Ta(2.0) multilayer. This trend is consistent with the current-induced magnetization switching measurements obtained via the anomalous Hall effect (see Fig. S5 in the supporting information). The key structural difference between the two stacks is the replacement of 5 nm Ta with Ti, suggesting that the OHE in the Ti layer plays a dominant role in reducing $J_c$ (Fig. 2d). In the Ta-only structure, magnetization switching is primarily driven by the spin current generated via the spin Hall effect in Ta. However, with the introduction of a Ti layer, the orbital current generated by the OHE in Ti is converted into spin current by the adjacent 0.5 nm thick Ta layer. This spin current then exerts a damping-like torque on the magnetization, leading to magnetization switching. We also characterize the damping-like torque efficiency ($\xi_{\mathrm{DL}}$)[41, 54-57] (see Figs. S6 to S8 in the Supporting information) of the Ti($t_{\mathrm{Ti}}$)/Ta(0.5)/CFB(0.75)/ MgO(2.2)/Ta(2) multilayers. The measured $\xi_{\mathrm{DLT}}$ within the range of 0.01~0.015 is smaller than the value of 0.045 for the pure Ta layer. The lower $\xi_{\mathrm{DLT}}$ may be due to the relatively low orbital to spin conversion efficiency of $\eta_{L-S}$ of ~0.012 of the Ta layer (see Fig. S8 in the Supporting information).

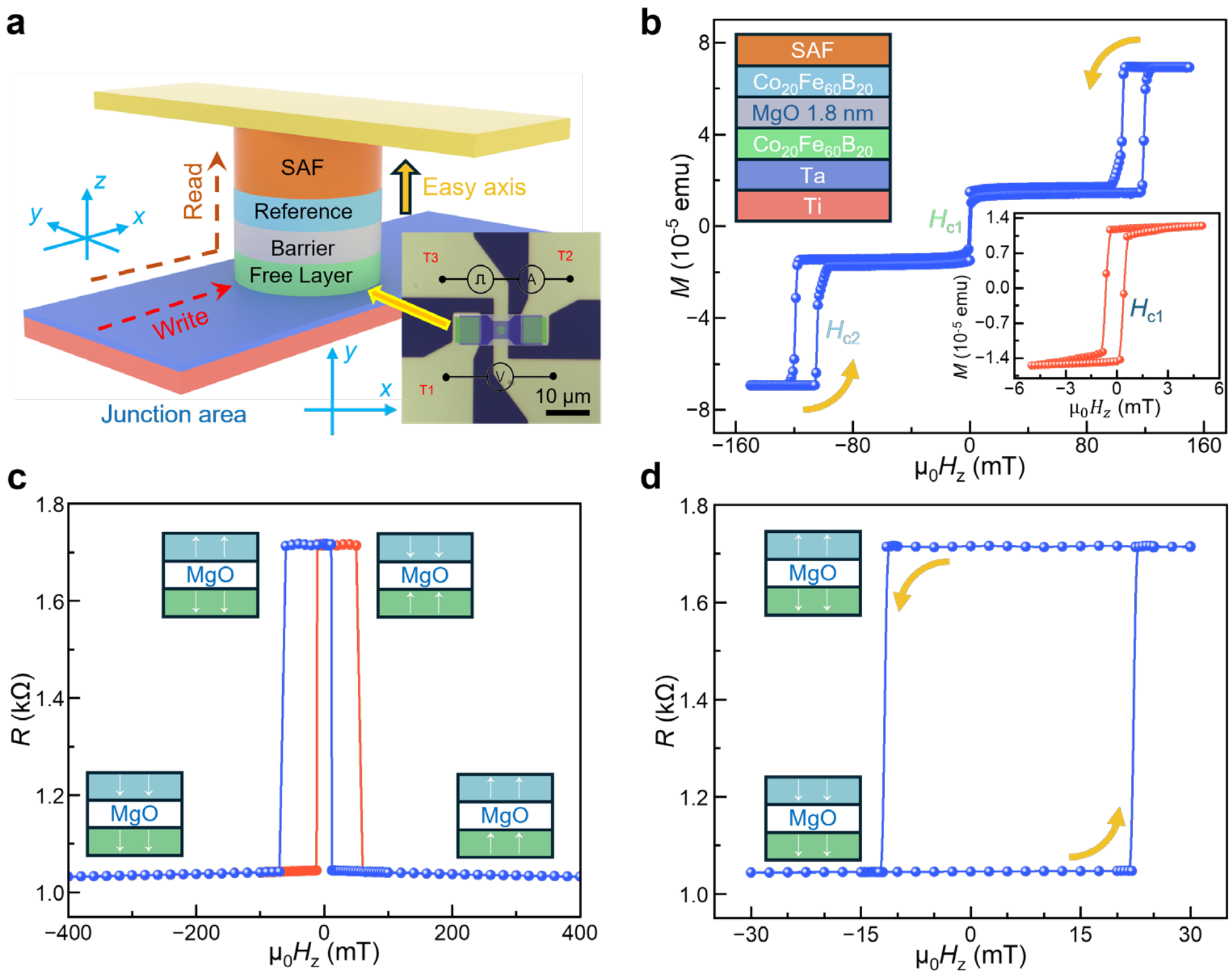

**Figure 3.** (a) Microscopic image and the schematic illustration of the patterned MTJ device. The Ti/Ta layer serves as the bottom electrode with a stripe width of 5 μm, and the radius of the MTJ pillar is 1 μm. (b) *M-H* loop of the multilayers measured at room temperature. The insects schematically illustrate the multilayer structure and the minor loop measurement of the free layer's magnetization switching. (c) TMR as a function of the magnetic field $\mu_0 H_z$. (d) Minor loop measurement of the TMR as a function of the magnetic field $\mu_0 H_z$.

## MTJ measurements

After demonstrating orbital current-induced magnetization switching in the Ti(5.0)/Ta(0.5)/CFB(0.75)/MgO(2.2)/Ta(2.0) multilayer, we performed current-induced magnetization switching in the pMTJs. The complete pMTJ device consists of the following stack sequence: Ti(5.0)/Ta(0.5)/CFB(0.75)/MgO(1.8)/CFB(1.05)/

Ta(0.5)/Co(0.3)/Pt(1.3)/Co(0.4)/Pt(0.6)/Co(0.4)/Ru(0.85)/[Co(0.4)/Pt(0.6)]$_4$/Ru(5.0). In this structure, the Ti and Ta layers serve as the bottom electrode. As shown in Fig. 3a, the bottom CFB layer (B-CFB), located directly above the bottom electrode, functions as the free layer of the MTJ; while the top CFB layer (T-CFB), located above the MgO barrier, acts as the reference layer. The magnetization of the reference layer is pinned by the SAF structure above the T-CFB layer. All layers are deposited on the oxidized Si(100) substrates at room temperature. After deposition, the stack is annealed at 300°C to improve the crystalline quality of the MgO barrier. During annealing, a perpendicular magnetic field of 800 mT is applied to establish a perpendicular magnetic easy axis (EA) for both the free and reference layers.[58] The magnetic hysteresis (*M-H*) loop (Fig. 3b), measured using a Magnetic Properties Measurement System (MPMS), is obtained by scanning the applied magnetic field along the EA of the multilayer stack. The measurement reveals a 2-step switching process, corresponding to distinct coercive fields for the B-CFB and T-CFB layers. This supports the presence of an antiparallel state between $H_{c1}$ (105 mT) and $H_{c2}$ (0.8 mT). (See Fig. S10 in supporting information for the detailed discussion of the *M-H* loop). Subsequently, the stack is patterned into MTJ devices using photolithography and Ar ion milling processes.

The magnetic field-driven magnetization switching of the MTJ is investigated by measuring the tunneling resistance across the top and bottom electrodes while varying the external magnetic field $\mu_0 H_z$ at room temperature. Fig. 3c shows the tunneling resistance $R$ as a function of the magnetic field, sweeping from 400 mT to -400 mT. The observed two-step switching in tunneling resistance aligns well with the trends observed in the *M-H* loop measured before device fabrication. From this switching behavior, a TMR ratio of 66% is achieved. Additionally, to confirm that the reference layer remains unchanged, a narrower range field scan is performed (Fig. 3d). This measurement distinctly captures the field-driven switching of the free layer, verifying the successful fabrication and proper functionality of the pMTJs.

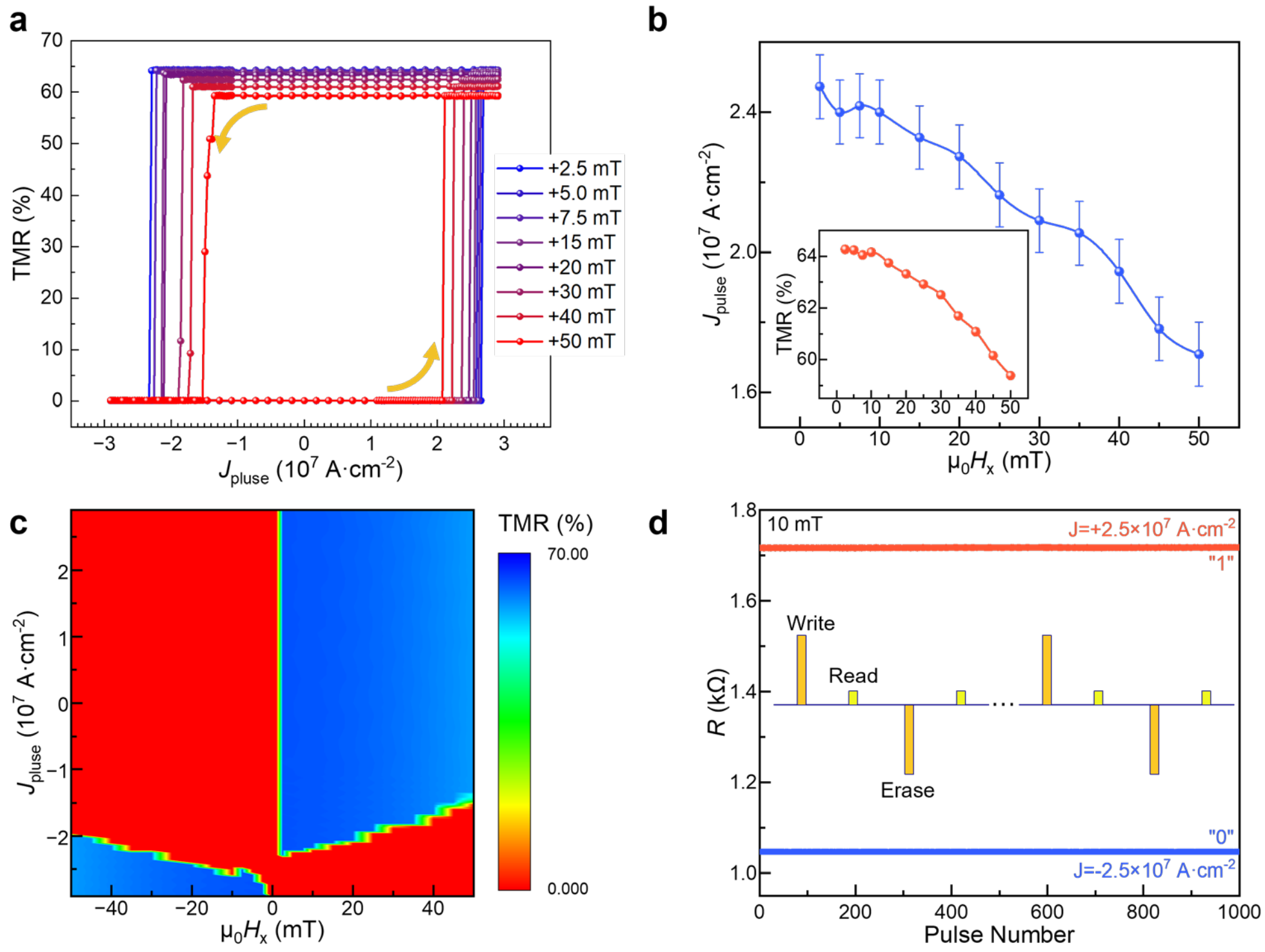

**Figure 4.** (a) Current-driven magnetization switching of the MTJ measured under different external magnetic fields $\mu_0 H_x$. (b) Critical switching current density $J_c$ as a function of $\mu_0 H_x$. The inset shows the TMR as a function of $\mu_0 H_x$. (c) Phase diagram of TMR ratio as a function of the pulse current density $J_c$ and external magnetic field $\mu_0 H_x$. (d) Endurance of the MTJ devices written by alternative positive and negative pulse currents. After each writing process, the tunneling resistance is measured by a reading current of 2.0 μA.

We then performed the current-driven magnetization switching measurements in the MTJ devices. During the measurement, both the pulse current and external magnetic field $\mu_0 H_x$ are applied along the *x*-axis. Prior to the switching measurement, the magnetization of the reference layer is preset in the up direction. As shown in Fig. 4a, successful current-induced magnetization switching has been achieved in the MTJ devices. A positive pulse current favors the high TMR ratio, while the negative pulse current favors the low TMR ratio, indicating antiparallel and parallel alignments of the magnetizations of the free and reference layers, respectively. Furthermore, the critical switching current densities for positive and negative pulses exhibit asymmetric behavior. At $\mu_0 H_x$ = 2.5 mT, the positive critical switching current density $J_{c,positive}$ is ~$2.7 \times 10^7$ A/cm$^2$, while the negative critical switching current density $J_{c,negative}$ is ~$2.2 \times 10^7$ A/cm$^2$. This asymmetry arises from the intrinsic preference for parallel magnetization alignment between the free and reference layers (see Fig. S12 in supporting information). Additionally, the average critical switching current density, defined as $J_c = (J_{c,positive}+J_{c,negative})/2$, is strongly influenced by the external field $\mu_0 H_x$, which decreases with increasing $\mu_0 H_x$, as is shown in Fig. 4b. This observation is consistent with the similar trend observed in the Ti/Ta/CFB/MgO/Ta multilayer shown in Fig. 2b. Moreover, the TMR ratio also decreases with increasing $\mu_0 H_x$, which can be attributed to the canting of perpendicular magnetization induced by the applied in-plane field $\mu_0 H_x$.

To gain a comprehensive understanding of the current driven magnetization switching in the MTJ devices, we conducted a phase diagram mapping measurement of the TMR ratio as a function of the pulse current density $J_c$ and external magnetic field $\mu_0 H_x$, as shown in Fig. 4c. In this diagram, the blue, red and green regions correspond to the high TMR ratio (antiparallel state), low TMR ratio (parallel state), and intermedium states, respectively. The following phenomena have been observed: First, the transition between the high and low TMR states occurs with narrow intermediate (green) regions, indicating a coherent magnetization switching process. This aligns well with the sharp switching behavior observed in Fig. 4a. Second, consistent with Fig. 4b, the critical switching current density, represented by the narrow green region, decreases as $\mu_0 H_x$ increases, further confirming the influence of the in-plane field $\mu_0 H_x$ on the switching dynamics. Finally, the magnetization switching polarity is found to be influenced by the sign of $\mu_0 H_x$. When scanning pulse current from positive to negative, the TMR ratio switches from high to low at positive $\mu_0 H_x$, whereas at negative $\mu_0 H_x$, the TMR ratio switches from low to high.

After successfully demonstrating orbital current-driven magnetization switching in pMTJ devices, we further evaluate their endurance by repeatedly switching between high- and low-resistance states. The test is conducted at room temperature under a constant in-plane magnetic field of 10 mT along the *x*-axis. Under these conditions, current pulses of $2.5\times10^7$ A/cm$^2$ are alternately applied through the bottom electrode in both +*x* and -*x* directions. After each writing pulse, the tunneling resistance is measured, consistently exhibiting two distinct resistance states, confirming stable and repeatable switching. Specifically, when a positive current is applied, the resistance increases to 1.72 kΩ, while a negative current reduces it to 1.05 kΩ. This highly stable bistable resistance behavior underscores the exceptional reliability of our pMTJ device, demonstrating its robust performance under continuous writing and erasing cycles.

In summary, we have successfully demonstrated orbital current-induced magnetization switching in Ti/Ta/CFB/MgO/CFB/SAF perpendicular magnetic tunnel junctions, achieving a high TMR ratio of 66%. The orbital torque-driven pMTJ exhibits stable performance, showing no signs of fatigue even after 1000 switching times. These findings highlight the potential of orbital torque-driven magnetization switching, for energy-efficient and reliable spintronic memory applications.

**Acknowledgments**

This work is supported by the King Abdullah University of Science and Technology, Office of Sponsored Research (OSR), under award Nos. ORA-CRG10-2021-4665, and ORA-CRG11-2022-5031. A.C. acknowledges support by the National Key Research and Development Program of China (No. 2024YFA1408503) and Sichuan Province Science and Technology Support Program (No. 2025YFHZ0147). Z.Q.Q. acknowledge support from US Department of Energy, Office of Science, Office of Basic Energy Sciences, Materials Sciences and Engineering Division under Contract No. DE-AC02-05CH11231 (van der Waals heterostructures program, KCWF16), Future Materials Discovery Program through the National Research Foundation of Korea (No. 2015M3D1A1070467), and Science Research Center Program through the National Research Foundation of Korea (No. 2015R1A5A1009962).

## Conflict of Interest

The authors declare that they have no conflict of interest.

## Reference

1. D. Apalkov, B. Dieny and J. M. Slaughter, Proceedings of the IEEE **104** (10), 1796-1830 (2016).
2. S. Bhatti, R. Sbiaa, A. Hirohata, H. Ohno, S. Fukami and S. N. Piramanayagam, Materials Today **20** (9), 530-548 (2017).
3. H. Ohno, M. D. Stiles and B. Dieny, Proceedings of the IEEE **104** (10), 1782-1786 (2016).
4. W. H. Butler, X. G. Zhang, T. C. Schulthess and J. M. MacLaren, Physical Review B **63** (5), 054416 (2001).
5. M. Julliere, Physics Letters A **54** (3), 225-226 (1975).
6. T. Miyazaki and N. Tezuka, Journal of Magnetism and Magnetic Materials **139** (3), L231-L234 (1995).
7. J. S. Moodera, L. R. Kinder, T. M. Wong and R. Meservey, Physical Review Letters **74** (16), 3273-3276 (1995).
8. S. S. P. Parkin, C. Kaiser, A. Panchula, P. M. Rice, B. Hughes, M. Samant and S.-H. Yang, Nature Materials **3** (12), 862-867 (2004).
9. S. Yuasa, T. Nagahama, A. Fukushima, Y. Suzuki and K. Ando, Nature Materials **3** (12), 868-871 (2004).
10. N. Sato, F. Xue, R. M. White, C. Bi and S. X. Wang, Nat. Electron. **1** (9), 508-511 (2018).
11. M. Wang, W. Cai, D. Zhu, Z. Wang, J. Kan, Z. Zhao, K. Cao, Z. Wang, Y. Zhang, T. Zhang, C. Park, J.-P. Wang, A. Fert and W. Zhao, Nat. Electron. **1** (11), 582-588 (2018).
12. H. Wu, P. Zhang, P. Deng, Q. Lan, Q. Pan, S. A. Razavi, X. Che, L. Huang, B. Dai, K. Wong, X. Han and K. L. Wang, Phys. Rev. Lett. **123** (20), 207205 (2019).
13. M. Dc, R. Grassi, J. Y. Chen, M. Jamali, D. Reifsnyder Hickey, D. Zhang, Z. Zhao, H. Li, P. Quarterman, Y. Lv, M. Li, A. Manchon, K. A. Mkhoyan, T. Low and J. P. Wang, Nat. Mater. **17** (9), 800-807 (2018).
14. Y. Shiomi, K. Nomura, Y. Kajiwara, K. Eto, M. Novak, K. Segawa, Y. Ando and E. Saitoh, Phys. Rev. Lett. **113** (19), 196601 (2014).
15. A. R. Mellnik, J. S. Lee, A. Richardella, J. L. Grab, P. J. Mintun, M. H. Fischer, A. Vaezi, A. Manchon, E. A. Kim, N. Samarth and D. C. Ralph, Nature **511** (7510), 449-451 (2014).
16. Y. Wang, D. Zhu, Y. Yang, K. Lee, R. Mishra, G. Go, S.-H. Oh, D.-H. Kim, K. Cai and E. Liu, Science **366** (6469), 1125-1128 (2019).
17. D. MacNeill, G. M. Stiehl, M. H. D. Guimaraes, R. A. Buhrman, J. Park and D. C. Ralph, Nat. Phys. **13** (3), 300-305 (2016).
18. X. Wang, J. Tang, X. Xia, C. He, J. Zhang, Y. Liu, C. Wan, C. Fang, C. Guo, W. Yang, Y. Guang, X. Zhang, H. Xu, J. Wei, M. Liao, X. Lu, J. Feng, X. Li, Y. Peng, H. Wei, R. Yang, D. Shi, X. Zhang, Z. Han, Z. Zhang, G. Zhang, G. Yu and X. Han, Sci. Adv. **5** (8) (2019).
19. H. Wang, K.-Y. Meng, P. Zhang, J. T. Hou, J. Finley, J. Han, F. Yang and L. Liu, Appl. Phys. Lett. **114** (23), 232406 (2019).
20. L. Liu, Q. Qin, W. Lin, C. Li, Q. Xie, S. He, X. Shu, C. Zhou, Z. Lim, J. Yu, W. Lu, M. Li, X. Yan, S. J. Pennycook and J. Chen, Nat. Nanotechnol. **14** (10), 939-944 (2019).

21. D. Zheng, J. Lan, B. Fang, Y. Li, C. Liu, J. O. Ledesma-Martin, Y. Wen, P. Li, C. Zhang, Y. Ma, Z. Qiu, K. Liu, A. Manchon and X. Zhang, Adv. Mater. **34** (34), 2203038 (2022).
22. D. Zheng, J. Xu, Q. Wang, C. Liu, T. Yang, A. Chen, Y. Li, M. Tang, M. Chen, H. Algaidi, C. Jin, K. Liu, M. Klaui, U. Schwingenschlogl and X. Zhang, Nano Lett. **25** (4), 1528-1535 (2025).
23. D. Zheng, Y. Li, C. Liu, J. Lan, C. Jin, Q. Wang, L. Zhang, G. Xi, B. Fang, C. Zhang, H. Algaidi, A. Chen, X. Liu, G. Yin, Z. Xu, J. Q. Xiao, A. Manchon and X. Zhang, Matter **7** (10), 3489-3499 (2024).
24. B. A. Bernevig, T. L. Hughes and S.-C. Zhang, Physical Review Letters **95** (6), 066601 (2005).
25. S. Ding, A. Ross, D. Go, L. Baldrati, Z. Ren, F. Freimuth, S. Becker, F. Kammerbauer, J. Yang, G. Jakob, Y. Mokrousov and M. Kläui, Physical Review Letters **125** (17), 177201 (2020).
26. D. Go, D. Jo, H.-W. Lee, M. Kläui and Y. Mokrousov, Europhysics Letters **135** (3), 37001 (2021).
27. S. Ding, Z. Liang, D. Go, C. Yun, M. Xue, Z. Liu, S. Becker, W. Yang, H. Du, C. Wang, Y. Yang, G. Jakob, M. Kläui, Y. Mokrousov and J. Yang, Physical Review Letters **128** (6), 067201 (2022).
28. Y.-G. Choi, D. Jo, K.-H. Ko, D. Go, K.-H. Kim, H. G. Park, C. Kim, B.-C. Min, G.-M. Choi and H.-W. Lee, Nature **619** (7968), 52-56 (2023).
29. R. Fukunaga, S. Haku, H. Hayashi and K. Ando, Physical Review Research **5** (2), 023054 (2023).
30. T. P. Cysne, L. M. Canonico, M. Costa, R. Muniz and T. G. Rappoport, arXiv preprint arXiv:2502.12339 (2025).
31. Z. Chen, R. Li, Y. Bai, N. Mao, M. Zeer, D. Go, Y. Dai, B. Huang, Y. Mokrousov and C. Niu, Nano letters **24** (16), 4826-4833 (2024).
32. T. P. Cysne, W. J. d. M. Kort-Kamp and T. G. Rappoport, Physical Review Research **6** (2), 023271 (2024).
33. D. Go, H.-W. Lee, P. M. Oppeneer, S. Blügel and Y. Mokrousov, Physical Review B **109** (17), 174435 (2024).
34. L. Salemi and P. M. Oppeneer, Physical Review Materials **6** (9), 095001 (2022).
35. S. Ding, A. Ross, D. Go, L. Baldrati, Z. Ren, F. Freimuth, S. Becker, F. Kammerbauer, J. Yang, G. Jakob, Y. Mokrousov and M. Klaui, Phys. Rev. Lett. **125** (17), 177201 (2020).
36. H. Hayashi, D. Jo, D. Go, T. Gao, S. Haku, Y. Mokrousov, H.-W. Lee and K. Ando, Communications Physics **6** (1), 32 (2023).
37. G. Sala and P. Gambardella, Phys. Rev. Res. **4** (3) (2022).
38. S. Lee, M.-G. Kang, D. Go, D. Kim, J.-H. Kang, T. Lee, G.-H. Lee, J. Kang, N. J. Lee, Y. Mokrousov, S. Kim, K.-J. Kim, K.-J. Lee and B.-G. Park, Commun. Phys. **4** (1) (2021).
39. D. Lee, D. Go, H. J. Park, W. Jeong, H. W. Ko, D. Yun, D. Jo, S. Lee, G. Go, J. H. Oh, K. J. Kim, B. G. Park, B. C. Min, H. C. Koo, H. W. Lee, O. Lee and K. J. Lee, Nat. Commun. **12** (1), 6710 (2021).
40. F. Liu, B. Liang, J. Xu, C. Jia and C. Jiang, Phys. Rev. B **107** (5) (2023).
41. R. Gupta, C. Bouard, F. Kammerbauer, J. O. Ledesma-Martin, A. Bose, I. Kononenko, S. Martin, P. Usé, G. Jakob, M. Drouard and M. Kläui, Nature Communications **16** (1), 130 (2025).
42. J. H. Cullen, H. Liu and D. Culcer, arXiv preprint arXiv:2501.17919 (2025).
43. D. Go and H.-W. Lee, Phys. Rev. Res. **2** (1), 013177 (2020).
44. D. Go, D. Jo, K. W. Kim, S. Lee, M. G. Kang, B. G. Park, S. Blugel, H. W. Lee and Y. Mokrousov, Phys. Rev. Lett. **130** (24), 246701 (2023).
45. Y. Yang, P. Wang, J. Chen, D. Zhang, C. Pan, S. Hu, T. Wang, W. Yue, C. Chen, W. Jiang, L. Zhu, X. Qiu, Y. Yao, Y. Li, W. Wang and Y. Jiang, Nat. Commun. **15** (1), 8645 (2024).
46. T. Li, L. Liu, X. Li, X. Zhao, H. An and K. Ando, Nano Lett. **23** (15), 7174-7179 (2023).
47. A. Bose, F. Kammerbauer, R. Gupta, D. Go, Y. Mokrousov, G. Jakob and M. Kläui, Phys. Rev. B

**107** (13), 134423 (2023).
48. J.-Y. Chen, Y.-C. Lau, J. M. D. Coey, M. Li and J.-P. Wang, Scientific Reports **7** (1), 42001 (2017).
49. S. Ikeda, K. Miura, H. Yamamoto, K. Mizunuma, H. D. Gan, M. Endo, S. Kanai, J. Hayakawa, F. Matsukura and H. Ohno, Nature Materials **9** (9), 721-724 (2010).
50. Q. Yang, L. Wang, Z. Zhou, L. Wang, Y. Zhang, S. Zhao, G. Dong, Y. Cheng, T. Min, Z. Hu, W. Chen, K. Xia and M. Liu, Nature Communications **9** (1), 991 (2018).
51. C. Muehlenhoff, M. Krupinski, A. Zarzycki and M. Albrecht, IEEE Sensors Journal **22** (6), 5588-5595 (2022).
52. Y. Liu, J. Yu and H. Zhong, Journal of Magnetism and Magnetic Materials **473**, 381-386 (2019).
53. K.-S. Lee, S.-W. Lee, B.-C. Min and K.-J. Lee, Appl. Phys. Lett. **102** (11), 112410 (2013).
54. S. y. Shin, D. Han, S. Lee and B.-G. Park, Advanced Functional Materials **n/a** (n/a), 2425932 (2025).
55. J. Kim, J. Sinha, M. Hayashi, M. Yamanouchi, S. Fukami, T. Suzuki, S. Mitani and H. Ohno, Nature Materials **12** (3), 240-245 (2013).
56. L. Liu, Q. Qin, W. Lin, C. Li, Q. Xie, S. He, X. Shu, C. Zhou, Z. Lim, J. Yu, W. Lu, M. Li, X. Yan, S. J. Pennycook and J. Chen, Nature Nanotechnology **14** (10), 939-944 (2019).
57. Y. Yang, P. Wang, J. Chen, D. Zhang, C. Pan, S. Hu, T. Wang, W. Yue, C. Chen, W. Jiang, L. Zhu, X. Qiu, Y. Yao, Y. Li, W. Wang and Y. Jiang, Nature Communications **15** (1), 8645 (2024).
58. S. Liang, A. Chen, L. Han, H. Bai, C. Chen, L. Huang, M. Ma, F. Pan, X. Zhang and C. Song, Advanced Functional Materials **n/a** (n/a), 2417731 (2024).